\documentclass{elsart}

\usepackage{graphicx}% Include figure files

\begin{document}

\begin{frontmatter}

\title{Subordination scenario of anomalous relaxation}

\author[label1]{Aleksander Stanislavsky\corauthref{cor1}}
\author[label2]{, Karina Weron\corauthref{cor2}}
\corauth[cor1]{\it{E-mail address}: \rm{alexstan@ira.kharkov.ua}}
\corauth[cor2]{\it{E-mail address}: \rm{Karina.Weron@pwr.wroc.pl}}
\address[label1]{Institute of Radio Astronomy, Ukrainian National Academy of
Sciences,\\ 4 Chervonopraporna St., 61002 Kharkov, Ukraine}
\address[label2]{Institute of Physics,  Wroc{\l}aw University of Technology,\\ Wyb.
Wyspia$\acute{n}$kiego 27, 50-370 Wroc{\l}aw, Poland}

\begin{abstract}
The non-exponential relaxation is shown to result from subordination by inverse
tempered $\alpha$-stable processes. The main feature of tempered $\alpha$-stable processes is a 
finiteness of their moments, and the class of random processes includes ordinary $\alpha$-stable processes as a 
particular case. Starting with the Cole-Cole response this subordination approach establishes its direct 
link with the Cole-Davidson law. We derive the relaxation function describing the tempered 
relaxation. The meaning of the empirical response function is clarified.
\end{abstract}


\end{frontmatter}

\section{Introduction}
The major feature of dynamical processes in many complex relaxing systems is their
stochastic background \cite{1,2}. Particularly, in any dielectric (complex) system under
an week external electric field (external action) only a part (active dipoles or objects)
of the total number of dipoles is directly governed by changes of the field. But even
those dipoles, not contributing to the relaxation dynamics, can have an effect on the
behavior of active dipoles \cite{3}. If the dipoles interact with each other, then their
evolution has a random character. Consequently, the behavior of such a relaxing system as
a whole will not be exponential in nature. In this case the macroscopic behavior of the
complex systems is governed by ``averaging principles'' like the law of large numbers
following from the theory of probability \cite{4}. The macroscopic dynamics of complex
systems is not attributed to any particular object taken from those forming the systems.
The finding out an ``averaged'' object representity for the entire relaxing system is
not simple. The relation between the local random characteristics of complex systems and
the universal deterministic empirical laws requires a probabilistic justification. There
are some points of view on this problem. One of well-known them is based on randomizing the
parameters of distributions that describes the relaxation rates in disordered systems.
With regard to the dielectric relaxation, each individual dipole in a complex system
relaxes exponentially, but their relaxation rates are different and obey a probability
distribution (continuous function) \cite{3,5}. This approach is successive for getting many
empirical response laws and their classification, but it sometime becomes
enough complicated to interpret their interrelations and to derive macroscopic response equations..

In this paper we suggest an alternative approach to the analysis of non-exponential
relaxation. It is based on subordination of random processes. Recall that in 
the theory of anomalous diffusion the notation of subordination occupies one of the 
most important places (see, for example, \cite{5a} and references therein). So, a 
subordinated process $Y(U(t))$ is obtained by randomizing the time clock of a random 
(parent) process $Y(t)$ by means of a random process $U(t)$ called the directing process. 
The latter process is also often referred to as the randomized time or operational 
time \cite{5b}. In the common case the process $Y$ may be both random and deterministic 
in nature. The subordination of random processes is a starting point for the anomalous 
diffusion theory. 

We develop this approach to relaxation processes. It gives an efficient method 
for calculating the dynamical evaluating averages of the relaxation processes. In this 
connection Section~\ref{par2} is devoted a presentation of recent achievements of this 
method. Starting with the description of the two-state system evolution as a Markovian 
process, we develop the analysis on subordinated random processes. The processes differ 
from the Markovian ones by the temporal variable becoming random. In this context the 
Cole-Cole relaxation is an evident example. In Section~\ref{par3} we consider 
the tempered $\alpha$-stable processes. They overcome the infinite-moment difficulty of 
the usual (not tempered) $\alpha$-stable processes. As applied to the anomalous diffusion, the tempering 
gives a preserving the subdiffusive behavior for short times whereas for long times the 
diffusion is something like normal. Using the processes in Section~\ref{par4}, we develop a 
subordination scheme for the description of the tempered relaxation. Section~\ref{par5} formulates major properties 
of such relaxation. We show that it has a direct relation to the well-known experimental laws 
of relaxation, in particular, to the Cole-Davidson law. Finally, the conclusions are drawn in Section~\ref{par6}.

\section{Relaxation in Two-State Systems}\label{par2}
The simplest ordinary interpretation of relaxation phenomena is based on the concept of a
system of independent exponentially relaxing objects (for example, dipoles) with different 
(independent) relaxation rates \cite{6}. The relaxation process, following this law (called 
Debye's), may be represented by behavior of a two-state system. Let $N$ be the common number
of dipoles in a dielectric system. If $N_\uparrow$ is the number of dipoles in the state
$\uparrow$, $N_\downarrow$ is the number of dipoles in the state $\downarrow$ so that
$N=N_\uparrow+N_\downarrow$. Assume that for $t=0$ the system is stated in order so that
the states $\uparrow$ dominate, namely
\begin{displaymath}
\frac{N_\uparrow(t=0)}{N}=n_\uparrow(0)=1,\quad
\frac{N_\downarrow(t=0)}{N}=n_\downarrow(0)=0\,,
\end{displaymath}
where $n_\uparrow$ is the part of dipoles in the state $\uparrow$, $n_\downarrow$ the
part in the state $\downarrow$. Denote the transition rate by $w$ defined from
microscopic properties of the system (for instance, according to the given Hamiltonian of
interaction and the Fermi's golden rule). In the simplest case (D relaxation) the kinetic 
equation takes the form
\begin{equation}
\cases{\dot n_\uparrow(t)-w\,\{n_\downarrow(t)-
n_\uparrow(t)\}=0,&\cr \dot n_\downarrow(t)-w\,\{n_\uparrow(t)-
n_\downarrow(t)\}=0,&\cr}\label{eqa1}
\end{equation}
where, as usual, the dotted symbol  means the first-order
derivative. The relaxation function for the two-state system is
\begin{displaymath}
\phi_{\rm D}(t)=1-2n_\downarrow(t)=2n_\uparrow(t)-1=\exp(-2wt).
\end{displaymath}
It is easy see that the steady state of the system corresponds to equilibrium with
$n_\uparrow(\infty)=n_\downarrow(\infty)=1/2$. Clearly, its response has also an exponential
character. However, this happens to be the case of dipoles relaxing irrespective of
each other and of their environment. If the dipoles interact with their environment, and
the interaction is complex (or random), their contribution in relaxation already will not
result in any exponential delay.

Assume that the interaction of dipoles with environment is taken into account with the
aid of the temporal subordination. We will consider the evolution of the number of dipoles in the 
states $\downarrow$ and $\uparrow$. This are parent random processes in the sense of subordination. 
They may be subordinated by another random process with a probability density, say $p(\tau,t)$. 
If $n_\uparrow(\tau)$ and $n_\downarrow(\tau)$, taking from Eq.(\ref{eqa1}) as probability laws of 
the parent processes, depend now on a local time $\tau$, then the resulting $n_\uparrow(t)$ and 
$n_\downarrow(t)$ after the subordination is determined by the integral relation
\begin{displaymath}
n_\uparrow(t)=\int^\infty_0 n_\uparrow(\tau)\,p(\tau,t)\,d\tau\,, \quad
n_\downarrow(t)=\int^\infty_0 n_\downarrow(\tau)\,p(\tau,t)\,d\tau\,.
\end{displaymath}
If the directing process (a new time clock or stochastic
time arrow \cite{9}) is an inverse $\alpha$-stable process, its probability density has 
the following Laplace image
\begin{equation}
p^{S}(\tau, t)=\frac{1}{2\pi j}\int_{Br} e^{ut-\tau
u^\alpha}\,u^{\alpha-1}\,du=t^{-\alpha}F_\alpha(\tau/t^\alpha)\,,
\label{eqb1}
\end{equation}
where $Br$ denotes the Bromwich path. This probability density has a simple physical
interpretation. It determines the probability to be at the internal time (or so-called
operational time) $\tau$ on the real (physical) time $t$. The function $F_\alpha(z)$ can
be expanded as a Taylor series. Besides, it has the Fox' H-function representation
\begin{displaymath}
F_\alpha(z)=H^{10}_{11}\left(z\Bigg|{(1-\alpha,\alpha)\atop
(0,1)}\right)=\sum_{k=0}^\infty\frac{(-z)^k}{k!
\Gamma(1-\alpha(1+k))}\,,
\end{displaymath}
where $\Gamma(x)$ is the ordinary gamma function. In the theory of anomalous diffusion
the random process $S(t)$ is applied for the subordination of L$\acute{\rm e}$vy (or
Gaussian) random processes \cite{10}. The inverse $\alpha$-stable process accounts for the
amount of time that a walker does not participate in the motion process \cite{11}. If the
walker participated all time in the motion process, the internal time and the physical
(external) time would coincide.

As was shown in \cite{9,12,13}, the stochastic time arrow can be applied to the general
kinetic equation. Then the equation describing a two-state system takes the following
form
\begin{equation}
\cases{D^\alpha
n_\uparrow(t)-w\,\{n_\downarrow(t)-n_\uparrow(t)\}=0,&\cr
D^\alpha
n_\downarrow(t)-w\,\{n_\uparrow(t)-n_\downarrow(t)\}=0,&\cr}\qquad
0<\alpha\leq 1,\label{eqc1}
\end{equation}
where $D^\alpha$ is the $\alpha$-order fractional derivative with respect to time.
Here we use the Caputo derivative \cite{13a}, namely
\begin{displaymath}
D^\alpha
x(t)=\frac{1}{\Gamma(n-\alpha)}\int^t_0\frac{x^{(n)}(\tau)}
{(t-\tau)^{\alpha+1-n}}\,d\tau,\quad n-1<\alpha<n,
\end{displaymath}
where $x^{(n)}(t)$ means the $n$-derivative of $x(t)$. The
relaxation function for the two-state system is written as
\begin{displaymath}
\phi_{\rm
CC}(t)=1-2n_\downarrow(t)=2n_\uparrow(t)-1=E_\alpha(-2wt^\alpha),
\end{displaymath}
where $E_\alpha(z)=\sum_{n=0}^\infty z^n/\Gamma(1+n\alpha)$ is the one-parameter
Mittag-Leffler function \cite{14}. It is important to notice that the relaxation function
corresponds to the Cole-Cole (CC) law. With reference to the theory of subordination the
CC law shows that the dipoles tend to equilibrium via motion alternating with stops so
that the temporal intervals between them is random. The random values are governed by an
inverse $\alpha$-stable subordinator.

The evolution of $n_\uparrow(t)$ and $n_\downarrow$ in Eq. (\ref{eqb1}) can be connected
with the Mittag-Leffler distribution. If $Z_n$ denotes the sum of $n$ independent random
values with the Mittag-Leffler distribution, then the Laplace transform of
$n^{-1/\alpha}Z_n$ is $(1+s^\alpha/n)^{-n}$, which tends to $e^{-s^\alpha}$ as $n$ tends
to infinity. Following the arguments of Pillai \cite{15}, this supports an infinity
divisibility of the Mittag-Leffler distribution. Due to the power asymptotic form (long
tail) the distribution with parameter $\alpha$ is attracted to the stable distribution
with exponent $\alpha$, $0<\alpha<1$. The property of the Mittag-Leffler distribution
permits one to determine a stochastic process. The process (called Mittag-Leffler's)
arises of subordinating a stable process by a directing (generalized) gamma process 
\cite{15}. In this case the relaxation function has the form
\begin{equation}
\phi_{\rm
HN}(t)=1-\sum_{k=0}^\infty\frac{(-1)^k\Gamma(b+k)}{k!\Gamma(b)\Gamma(1+ab+ak)}
\left(t/\tau_{\rm HN}\right)^{ab+ak}\,, \label{eq13a}
\end{equation}
where $a$, $b$, $\tau_{\rm HN}$ are constant. The one-side Fourier
transformation of the relaxation function gives the Havriliak-Negami (HN) law
\begin{equation}
\chi_{\rm HN}(\omega)=\int^\infty_0e^{-i\omega t}\,\left(-\frac{d\phi_{\rm
HN}(t)}{dt}\right)\,dt= \frac{1}{(1+(i\omega\tau_{\rm HN})^a)^b}\,. \label{eqe13b}
\end{equation}
This result also corresponds to the well-know HN empirical law. Thus, the HN relaxation
can be explained from the subordination approach, if the hitting time process of dipole
orientations transforms into the Mittag-Leffler process \cite{16}. For that the hitting
time process has an appropriate distribution attracted to the stable distribution. The
subordination of the latter results just in the Mittag-Leffler process. It is interesting
to observe that the L\'evy process subordinated by another L\'evy one leads again to the
L\'evy process, but with other index \cite{17}. Unfortunately, the description from the 
Mittag-Leffler process gives nothing for the derivation of any macroscopic response equation
like Eq.(\ref{eqc1}).

\section{Tempered $\alpha$-stable Process and Its Inverse}\label{par3}
The relaxation model based on the inverse $\alpha$-stable process starts with the consideration
of $\alpha$-stable processes having the infinite-moment difficult. To overcome it, one can develop
an approach stated on tempered $\alpha$-stable processes. The tempered $\alpha$-stable process 
\cite{18,19} has the Laplace image of its distribution in the form
\begin{equation}
\tilde{f}(u)=\exp\left(\delta^\alpha-(u+\delta)^\alpha\right)
\,.\label{eq5}
\end{equation}
When $\delta$ equals to zero, the tempered $\alpha$-stable process
becomes simply $\alpha$-stable.

However, the distribution (\ref{eq5}) describes only probabilistic
properties in terms of internal time. For subordination we need the probability distribution
of the inverse tempered $\alpha$-stable process. If $f(\tau,t)$ is the p.d.f. of internal time, then the
p.d.f. of its inverse $g(\tau,t)$ can be represented as
\begin{displaymath}
g(\tau,t)=-\frac{\partial}{\partial\tau}\int_{-\infty}^tf(t',\tau)\,dt'.
\end{displaymath}
Taking the Laplace transform of $g(\tau,t)$ with respect to $t$,
we get
\begin{equation}
\tilde{g}(\tau,u)=-\frac{1}{u}\frac{\partial}{\partial\tau}\tilde{f}(u,\tau)=
\frac{(u+\delta)^\alpha-\delta^\alpha}{u}\,e^{-\tau[(u+\delta)^\alpha-\delta^\alpha]}
\,.\label{eq6}
\end{equation}
When $\delta\to 0$, Eq. (\ref{eq6}) tends to
\begin{displaymath}
\tilde{g}(\tau,u)=u^{\alpha-1}\,e^{-\tau u^\alpha}\,.
\end{displaymath}
This expression corresponds to the Laplace image of an inverse $\alpha$-stable p.d.f.
describing a directing process in the theory of Cole-Cole relaxation. After the inverse Laplace
transform we have Eq.(\ref{eqb1}). 

In the Laplace space the function $g(\tau,u)$ has a simple
form. Because of general properties of the Laplace transform we can find $g(\tau,t)$ explicitly, 
but its representation is enough complicated and expressed through a integral of the Wright functions \cite{19a}.
For our calculation it will be sufficient to know only the function $g(\tau,u)$. Therefore, we will not 
present any explicit form $g(\tau,t)$ here.

\section{Macroscopic Response Equation of Tempered Relaxation}\label{par4}
If in Eq.(\ref{eqa1}) the value $w$ will depend on time as $a\,A^at^{a-1}$, we come to the description
of the Kohlrausch-Williams-Watts (KWW) relaxation. Although such a equation does not contain any 
(for example, micro/mesoscopic and so on) details about relaxation processes, it is convenient for 
practical purpose because of its simplicity. When the relaxation follows the CC, CD (Cole-Davidson), 
HN laws, the equation (\ref{eqa1}) is not so simple as in the case of D and KWW relaxation. Recall
that the CC relaxation and response functions can be expressed in terms of a solution of
the fractional differential equation \cite{17}. With macroscopic equations for the CD and HN 
responses the situation becomes more else complicated. Consider the derivation of the macroscopic 
response equation of tempered relaxation in more details.

For a general type of a Markovian process the general kinetic
equation is
\begin{equation}
\frac{dp\,(t)}{dt}=\hat{\bf W}p\,(t)\,, \label{eq12}
\end{equation}
where $\hat{\bf W}$ denotes the transition rate operator (see details, in \cite{23}).
This equation defines the probability $p$ for the system transition from one state into
others. Next, we determine a new process governed by an inverse tempered $\alpha$-stable
process with the Laplace image (\ref{eq6}), namely
\begin{displaymath}
p_\alpha(t)=\int_0^\infty g(t,\tau)\,p(\tau)\,d\tau\,.
\end{displaymath}
The Laplace transform $\tilde p_\alpha(s)$ is given by
\begin{displaymath}
\tilde p_\alpha(s)=\int_0^\infty e^{-st}\,p_\alpha(t)\,dt
\end{displaymath}
and leads to
\begin{eqnarray}
\hat{\bf W}\,\tilde
p_\alpha(s)&=&\frac{(s+\delta)^\alpha-\delta^\alpha}{s}\,\hat{\bf
W}\,\tilde p\,((s+\delta)^\alpha-\delta^\alpha)\nonumber\\
&=&\frac{(s+\delta)^\alpha-\delta^\alpha}{s}\,\Big\{[(s+\delta)^\alpha-
\delta^\alpha]\,\tilde p\,((s+\delta)^\alpha-\delta^\alpha)-p\,(0)\Big\}\nonumber\\
&=&[(s+\delta)^\alpha-\delta^\alpha]\,\tilde
p_\alpha(s)-\frac{(s+\delta)^\alpha-\delta^\alpha}{s}\,p\,(0)\,.
\label{eq13}
\end{eqnarray}
From this it follows
%The tempered extension of Eq. (\ref{eq12}) can be written as
\begin{equation}
p_\alpha(t)=p\,(0)+\int^t_0d\tau\,M(t- \tau)\,\hat{\bf W}\,p_\alpha(t)\,, \label{eq14}
\end{equation}
where the kernel $M(t)$ is written as
\begin{displaymath}
M(t)=e^{-\delta t}\,t^{\alpha-1}\,E_{\alpha,\,\alpha}(\delta^\alpha t^\alpha)\,,
\end{displaymath}
where $E_{\alpha,\beta}(z)=\sum_{n=0}^\infty z^n/\Gamma(\beta+n\alpha)$ is the
two-parameter Mittag-Leffler function \cite{14}.
This equation covers a number of particular cases known earlier. For $\alpha=1$ we obtain 
Eq. (\ref{eq12}), and for $\delta=0$ it becomes fractional. If a system has discrete 
states, then the generating function is of the form
\begin{displaymath}
G(\zeta, t)=\sum_{k=0}^\infty\zeta^kp_k(t)\,,
\end{displaymath}
where $\zeta$ takes values $\mid\zeta\mid\leq 1$ for a series to
converge. With the help of the generating function, one can find
the moments by taking the derivative with respect to $\zeta$ and
then setting $\zeta=1$. The generating function  of the process
governed by the stochastic time clock is given by the relation
\begin{equation}
G_\alpha(\zeta, t)=\int^\infty_0g(\tau,t)\,G(\zeta,\tau)\,d\tau\,.
\label{eq15}
\end{equation}
Thus, the generating function for a discrete Markovian process directed by a subordinated
process can be obtained from the appropriate generating function of the parent process by
immediate integration.

\begin{figure}
\center
\includegraphics[width= 13 cm]{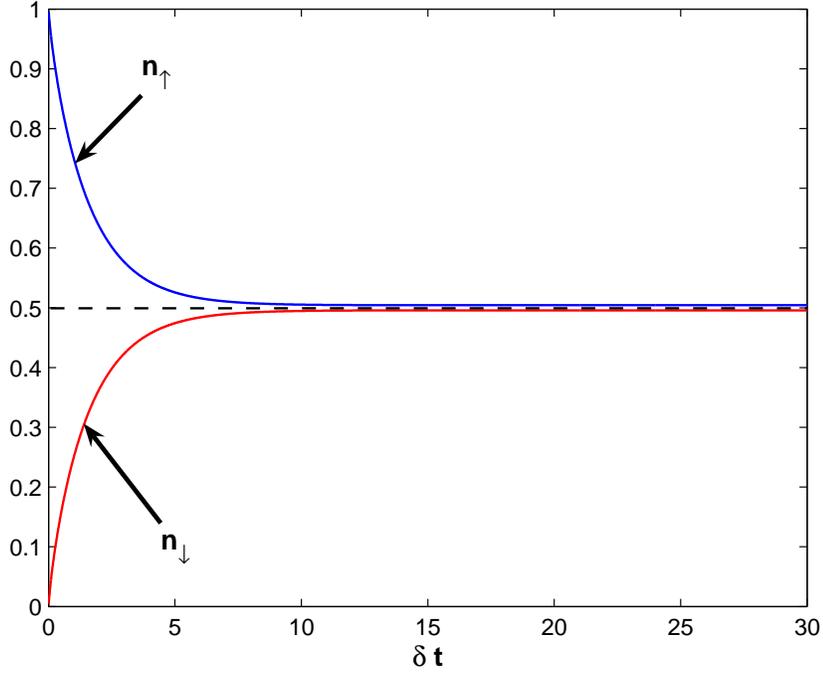}
\caption{\label{fig1}The relaxation of the part $n_\uparrow$ of dipoles in the state $\uparrow$ and 
the part $n_\downarrow$ of dipoles in the state $\downarrow$ (for $w/\delta$ = 1/4, $\alpha$ = 0.75).}
\end{figure}

\begin{figure}
\center
\includegraphics[width=13 cm]{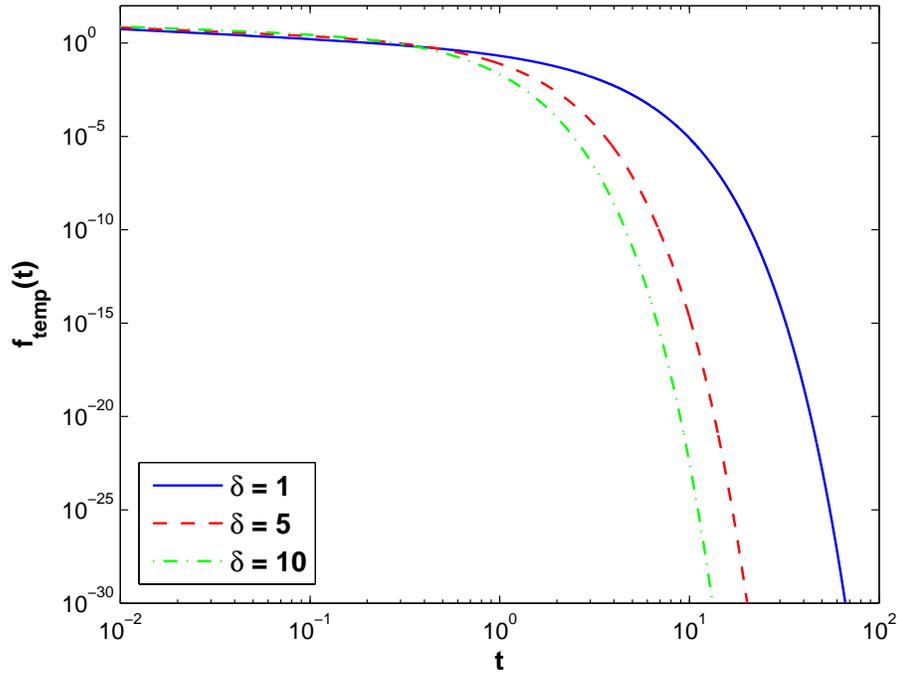}
\caption{\label{fig3}Response function with different $\delta$ ($w$ = 0.5, $\alpha$ = 0.5).}
\end{figure}

The relaxation in a two-state system under the inverse tempered $\alpha$-stable
subordinator gives
\begin{eqnarray}
n_\uparrow(t)&=&n_\uparrow(0)+w\int^t_0M(t-
\tau)\,\{n_\downarrow(t)-n_\uparrow(t)\}\,d\tau,\nonumber\\
n_\downarrow(t)&=&n_\downarrow(0)+w\int^t_0M(t-
\tau)\,\{n_\uparrow(t)-n_\downarrow(t)\}\,d\tau\,.\label{eq15a}
\end{eqnarray}
For $t\ll 1$ (or $\delta\to 0$) the
kernel $M(t)$ takes the power form $t^\alpha/\Gamma(\alpha)$ as a fractional kernel in the
integral representation of Eq. (\ref{eqc1}). However, for  $t\gg 1$ (or $\alpha\to 1$)
$M(t)$ becomes constant and, as a result, Eq. (\ref{eq15a}) transforms into the integral
form of the ordinary equations (\ref{eqa1}). From the linearity of these equations it just
follows
\begin{eqnarray}
n_\uparrow(t)+n_\downarrow(t)&=&n_\uparrow(0)+n_\downarrow(0)\,, \quad
n_\uparrow(t)-n_\downarrow(t)=\nonumber\\ n_\uparrow(0)-n_\downarrow(0)&-&2w
\int^t_0M(t-\tau)\,\{n_\uparrow(\tau)- n_\downarrow(\tau)\}\,d\tau.\nonumber
\end{eqnarray}
Consequently, we obtain
\begin{eqnarray}
n_\uparrow(t)&=&1-w\int_0^te^{-\delta\tau}\,\tau^{\alpha-1}\,E_{\alpha,\alpha}
([\delta^\alpha-2w]\tau^\alpha)\,d\tau,\nonumber\\
n_\downarrow(t)&=&w\int_0^te^{-\delta\tau}\,\tau^{\alpha-1}\,E_{\alpha,\alpha}
([\delta^\alpha-2w]\tau^\alpha)\,d\tau.\nonumber
\end{eqnarray}
The equations have steady states ($t\to\infty$) corresponding to equilibrium in this
system. According to \cite{24}, we know
\begin{displaymath}
\int^\infty_0e^{-ax}\,x^{\alpha-1}\,E_{\alpha,\alpha}(\pm bx)\,dx=\frac{1}{a^\alpha\mp
b},\quad ({\rm Re}(a)>|b|^{1/\alpha}),
\end{displaymath}
then $n_\uparrow(\infty)=n_\downarrow(\infty)=1/2$ for any $\delta\geq 0$ and $w>0$ (see,
for example, Fig.~\ref{fig1}). The transition rate $w$ is again defined by microscopic
properties of the system. Thus the relaxation function takes the form
\begin{displaymath}
\phi_{\rm temp}(t)=1-2n_\downarrow(t)=1-2w\int_0^t
e^{-\delta\tau}\,\tau^{\alpha-1}\,E_{\alpha,\alpha}
([\delta^\alpha-2w]\tau^\alpha)\,d\tau.
\end{displaymath}
The response function $f_{\rm temp}(t)=-d\phi_{\rm temp}(t)/dt$ is written as
\begin{displaymath}
f_{\rm temp}(t)=2w\,
e^{-\delta t}\,t^{\alpha-1}\,E_{\alpha,\alpha}
([\delta^\alpha-2w]t^\alpha)\,.
\end{displaymath}
Fig.~\ref{fig3} demonstrates how to change the response function with the increase of $\delta$.

\begin{figure}
\center
\includegraphics[width=13 cm]{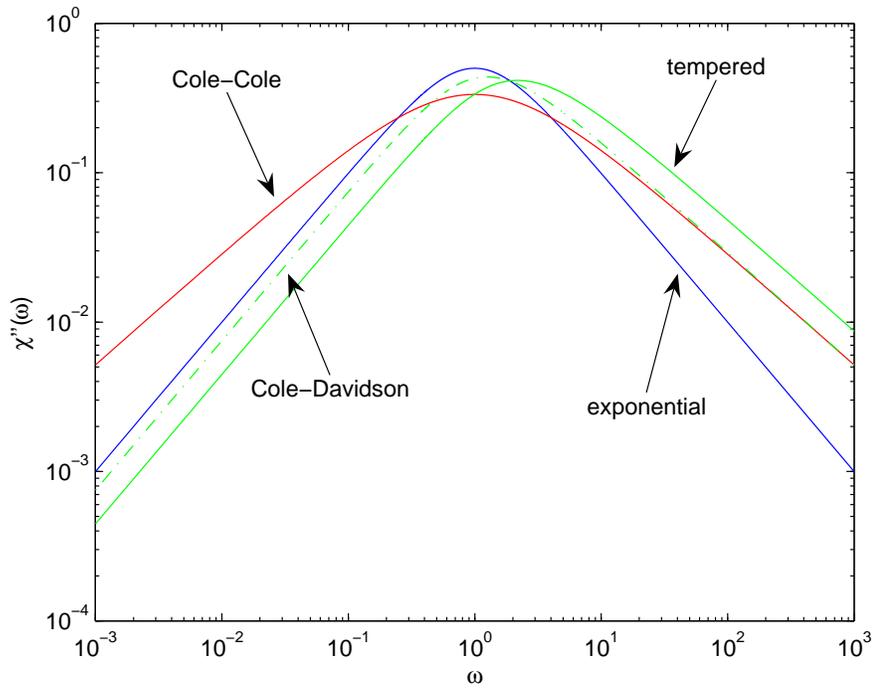}
\caption{\label{fig2}Imaginary term of the frequency-domain relaxation function
$\chi(\omega)=\chi'(\omega)-i\chi''(\omega)$.}
\end{figure}

For the experimental study it is interesting to get the frequency-domain representation of the latter
function. Its real part describes a dispersion of the relaxing medium, and its imaginary part is an absorption.
The values explicitly  can be measured in experiments. The one-side Fourier transformation of the response 
function gives
\begin{equation}
\chi_{\rm temp}(\omega)=\int^\infty_0e^{-i\omega t}\,\left(-\frac{d\phi_{\rm
temp}(t)}{dt}\right)\,dt= \frac{1}{1-\sigma^\alpha+(i\omega/\omega_{\rm
p}+\sigma)^\alpha}\,,\label{eqe16}
\end{equation}
where $\omega_p=(2w)^{1/\alpha}$ and $0\leq\sigma=\delta/(2w)^{1/\alpha}<\infty$ are
constant. The parameter $\omega_p$ is the characteristic frequency of the relaxing
system. It is easy to notice that the expression (\ref{eqe16}) for $\alpha=1$ is reduced
to the D law, for $\delta=0$ (or $\sigma=0$) it describes the CC relaxation, and for
$\sigma=1$ it does the CD law (see Fig.~\ref{fig2}).

\section{Major Properties of Tempered Relaxation}\label{par5}
The frequency dependence of the dielectric susceptibility for orientational polarization
of dipoles has been the subject of experimental and theoretical studies for many years,
but still there is no any generally accepted theory capable of explaining the observed
phenomena. Experimentally it is well known that the complex dielectric susceptibility
$\chi(\omega)=\chi'(\omega)-i\chi''(\omega)$ of most dipolar substances demonstrates a
peak in the loss component $\chi''(\omega)$ at a characteristic frequency $\omega_{\rm
p}$, and it is characterized by high- ($\omega\gg\omega_{\rm p}$) and low-frequency
($\omega\ll\omega_{\rm p}$) dependencies. The tempered relaxation shows
\begin{eqnarray}
\chi_{\rm temp}'(\omega)&=&\frac{A+B\cos(C)}{A^2+2AB\cos(C)+B^2}\,,\nonumber\\ \chi_{\rm
temp}''(\omega)&=&\frac{B\sin(C)}{A^2+2AB\cos(C)+B^2}\,,\nonumber
\end{eqnarray}
where $A=1-\sigma^\alpha$, $B=(\sigma^2+\omega^2/\omega_{\rm p}^2)^{\alpha/2}$ and
$C=\alpha\arctan(\omega/\delta)$. For small $\omega$ and any positive $\sigma\neq 0$ 
it is easy to see that
$\lim_{\omega\to 0}\chi_{\rm temp}'(\omega)\sim\omega$ and $\lim_{\omega\to 0}\chi_{\rm
temp}''(\omega)\sim 1$ whereas for large $\omega$ we get $\lim_{\omega\to\infty}\chi_{\rm
temp}'(\omega)\sim\omega^{-\alpha}$ and $\lim_{\omega\to\infty}\chi_{\rm
temp}''(\omega)\sim\omega^{-\alpha}$. This implies that
\begin{displaymath}
\lim_{\omega\to\infty}\frac{\chi_{\rm temp}''(\omega)}{\chi_{\rm
temp}'(\omega)}=\tan\Big(\frac{\alpha\pi}{2}\Big) =\cot\Big(n\frac{\pi}{2}\Big)\,,
\end{displaymath}
where $n=1-\alpha$, that is in agreement with the experimental results \cite{1,2}.
However, for small $\omega$ we come to
\begin{displaymath}
\lim_{\omega\to 0}\frac{\chi_{\rm temp}''(\omega)}{\chi_{\rm temp}'(0)-\chi_{\rm
temp}'(\omega)}=\infty\,.
\end{displaymath}
%while it should tend to $\tan(m\pi/2)$ with $0<m<1$, according to \cite{1}.
This means that the energy lost per cycle does not have a constant relationship to the
extra energy that can be stored by a static field. According to such an asymptotic
behavior and in Fig.~\ref{fig2} it is seen that the tempered relaxation takes an
intermediate place between the D, CC and CD types of relaxation.

%From (\ref{eqe16}) we get the only possible form
%\begin{displaymath}
%\phi_{\rm temp}(t)\approx\cases{ 1\,,\quad&if\quad $t\ll 1$;\cr 0\,,\quad&if\quad $t\gg
%1$.\cr} .
%\end{displaymath}

\section{Conclusions}\label{par6}
In this paper we have represented our progress in the subordination analysis of relaxation
phenomena in the complex systems. The approach permits ones to consider many relaxation
laws on the unique theoretical base originating from the stochastic nature of relaxation.
The general probabilistic formalism treats the relaxation of the complex systems
regardless of the precise nature of local interactions. Following this approach, we have
derived the empirical relaxation laws and their macroscopic equations, have characterized
their parameters, have connected the parameters with local random characteristics of the
relaxation processes, have demonstrated how to make the transition from the microscopic
random dynamics in the complex stochastic systems to the macroscopic deterministic
description by integro-differential equations. It should be pointed out that the form of
these equations is a direct sign of complexity evolution in such systems. Although we
restricted only by the detailed analysis of two-state systems, due to Eq.(\ref{eq14}), this 
consideration can be developed to the study of many-state systems (as an example, see the analysis of 
the three-state fractional system in \cite{13})). The tempered relaxation establishes a connection
between several types of relaxation (D, CC and CD). Its asymptotic behavior earnestly
shows that starting as a non-exponential relaxation, latter it tends to the D law.
Moreover, the theory of subordination suggests a clear interpretation of the tempered
relaxation. As applied to the dielectric relaxation, the interaction of dipoles with each
other and their environment has a confined time of action on the relaxation process near
a starting point of relaxation. Latter they behave independently just as this is the case
of exponential (D) relaxation. To put it in another way, the dipoles are strongly
connected with each other in the initial stage of relaxation (and, therefore, their
response function obeys a non-exponential decay), but the connection is not long-lived,
and in an interval of time each dipole evaluates on its own.

\section*{Acknowledgments}
AS is grateful to the Institute of Physics and the Hugo Steinhaus
Center for pleasant hospitality during his visit in Wroc{\l}aw University of Technology.

%\section*{References}

\end{document}